\documentclass[aps,twocolumn,longbibliography]{revtex4-1}
\usepackage{amssymb}
\usepackage[export]{adjustbox}
\usepackage{graphicx}
\usepackage[dvipsnames]{xcolor}
\usepackage{hyperref}
\begin{document}
\author{Shikha Bhadoria$^{1,2}$}\email{shikha.bhadoria@physics.gu.se}
\author{Mattias Marklund$^2$}
\author{Christoph H. Keitel$^1$} 
\affiliation{$^1$Max-Planck-Institut f\"ur Kernphysik, Saupfercheckweg 1, 69117 Heidelberg, Germany}
\affiliation{$^2$Department of Physics, University of Gothenburg, Sweden}

\title{Energy enhancement of laser-driven ions by radiation reaction and Breit-Wheeler pair production in the ultra-relativistic \textcolor{black}{transparency} regime} 

\begin{abstract}
The impact of radiation reaction and Breit-Wheeler pair production on acceleration of fully ionized Carbon ions driven by an intense linearly-polarized laser pulse has been investigated in the ultra-relativistic transparency regime. 
Against initial expectations radiation reaction and pair production at ultra-high laser intensities is found to enhance the energy gained by the ions. 
The electrons lose most of their transverse momentum and the additionally produced pair plasma of Breit-Wheeler electrons and positrons co-stream in the forward direction as opposed to the existing electrons streaming at an angle above zero. \textcolor{black}{ We discuss how these observations could be explained by the}
changes in the phase velocity of the Buneman instability, that is known to aid ion acceleration in the Breakout-Afterburner regime, by tapping the free energy in the relative electron and ion streams. We present evidence that \textcolor{black}{these non-classical effects} can further improve the highest Carbon ion energies \textcolor{black}{in this transparency regime}. 
\end{abstract}

\maketitle
\section{Introduction}
 Accelerated ion beams have a multitude of applications ranging from nuclear reactions induced by energetic heavy ions~\cite{McKenna2003} to fast ignition fusion~\cite{Roth2001,Fernandez_2014}, aiding neutron production~\cite{MRoth2013} and also hadrontherapy for cancer treatment~\cite{Bulanov2002,Schreiber2016, Bulanov_2014}. Laser-driven ion acceleration has acquired much attention in the recent decades, as this offers the possibility of having alternate accelerators that are smaller and more affordable as opposed to the conventional linacs, cyclotrons and synchrotron~\cite{Macchi2013a,Daido_2012}. Experimental demonstration of ion beams by several mechanisms exhibiting different performances such as Target Normal Sheath Acceleration (TNSA)~\cite{Snavely.2000}, Radiation- Pressure Accelerations (RPA)~\cite{Esirkepov.2004, Robinson2009, Palmer2012}, Collisionless shock acceleration (CSA)~\cite{Haberberger2012,Silva2004}, Breakout Afterburner (BOA)~\cite{Yin2007,Albright2007,Hegelich2013,Henig2009} etc. has already been achieved~\cite{Hilz2018}. Significant efforts of innovative laser/target configurations have also been made to push the numbers of ion beam characteristics (energies and flux)~\cite{Wan.2022}, yet the highest gained energy is still less than ~100 MeV/u~\cite{Higginson2018,Hilz2018,Keppler2022}. Nevertheless, the prospects of achieving even higher ion energies as predicted with the next generation laser sources are promising~\cite{Schreiber2006}. 
 
 BOA is one of the high performance laser-driven-ion acceleration mechanisms capable of accelerating ions to relatively higher values even with state-of-the-art lasers. In this, an initially opaque, ultra-thin target (width around laser skin depth) turns transparent to the incoming laser pulse, due to lowering of the density by the expanding plasma and increase in critical density by the electron's relativistic motion (relativistically induced transparency, RIT)~\cite{Yin2007,Henig2009a}. This leads to a phase of extreme ion acceleration (BOA phase) which continues to exist until the electron density of the expanding target becomes classically underdense~\cite{Yin2011a}. Buneman instability (in single ion-species target) and ion-ion acoustic instability (in case of multispecies target~\cite{King2017}) result in an electrostatic mode structure, that is found to be instrumental in transferring the laser energy to ions via laser-induced electronic drifts~\cite{Albright2007,King2016a}. The efficiency of this mechanism is maximised when the peak of the laser pulse arrives precisely at the onset of relativistic transparency~\cite{Hegelich2013,Petrov2017} 
 as opposed to the RPA-Light-sail mechanism which requires opacity in ultra-thin targets. 
 Experimental demonstration of fully ionized carbon ion acceleration via the BOA mechanism  up to 40-50 MeV/u has been achieved using $ \sim $50-250 nm thick targets with the TRIDENT laser and the Texas Pettawatt laser facility~\cite{Hegelich2013}. Also, simultaneously existing TNSA and BOA signatures in proton spectra (energy $\sim 61$MeV) have been identified at the PHELIX laser facility at GSI with 200-1200 nm targets with a 4-8$\times10^{22}$W/cm$^2$ laser~\cite{Wagner2015}. \textcolor{black}{Recently measured 30 MeV Carbon ions in the transparency regime are shown to be accelerated by extremely localised axial fields at the J-KAREN-P facility (also complemented by experiment at DRACO-PW)~\cite{Dover2023}}.
 Much more intense and powerful lasers, such as ELI, APOLLON, are soon to surface~\cite{papadopoulos2016,Bagnoud2010,Gales2018} (as expected in the laser-power timeline and also with the recent prototype design using WNOPCPA allowing a 0.5 EW system~\cite{Li2021}) and can further improve these numbers, as they will allow a larger laser energy transfer to the ions. However, in the ultra-reltivistic regime other Quantum Electrodynamic Dynamic (QED) effects become non-negligible when the electric field of the laser in the electron's rest frame gets closer to the critical Schwinger field ($E_s = 1.3826 \times 10^{18} V$m$^{-1}$~\cite{Schwinger1951}). The most important effects are: high frequency radiation emission by electrons pushed in the laser-field (with a consequent back reaction on individual electrons, radiation reaction (RR)) and the multi-photon Breit-Wheeler process leading to the generation of electron-positron pairs~\cite{BreitWheeler1934}. These QED effects, usually expected to deplete energy from a physical system~\cite{Bhadoria2019,Tamburini2010,Tamburini2011, wallin_gonoskov_harvey_lundh_marklund_2017,DelSorbo2018}, may though significantly modify the collective plasma dynamics~\cite{Capdessus_2020} with yet unexplored indirect effects on the ion energy.

In this paper, the impact of both RR and non-linear Breit-Wheeler pair production (PP, $\gamma+n\gamma\to e^-e^+$) on the acceleration of ions in the \textcolor{black}{transparency} regime has been investigated using PIC simulations. An increase in ion energies by RR alone in the transparency regime has already been reported~\cite{Tamburini2010,Tamburini2011,Tamburini2012,min.chen.iop.2010,Capdessus2015,Gelfer2021} , though these neglect the stochastic nature of high-energy photon emission. Here, we show evidence that in this regime, both RR and PP together can lead to a notable improvement upto $ 30\%$ in the ion energy beyond previous results. \textcolor{black}{This is attributed to more collimation of the plasma stream due to QED effects. Though, co-existence of a less-efficient RPA can not be ruled out~\cite{Bulanov.2016}, we present a discussion on how the observed spectra could also be explained via the BOA mechanism by identifying low-frequency electrostatic modes in the spectral analysis of the system. Then the improvement in ion energies are explained by an enhancement of the phase velocity of the relativistic Buneman instability (RBI) that is responsible to accelerate ions via Landau damping~\cite{Stark2018}. This allows for an efficient energy transfer from the laser to the ions facilitated by electron flow during the onset of RIT.}

\section{Simulations}\label{simulations}
We performed 2D PIC simulations using both the open-source codes EPOCH and SMILEI which include quantum RR and PP by the probabilistic Monte-Carlo method~\cite{Kirk2009,Duclous_2011}. We employ a linearly $s-$polarized laser pulse, impinging on the left boundary with a finite spatio-temporal profile $I(t,y)= I_0 \exp [-((y-y')/r_0)^2] \exp [-((t-t')/\tau_0)^2]$, with $r_0=3\, \mu$m, $y'=4\, \mu$m, $\tau_0=40$ fs, $t'=30$ fs. The laser peak intensity of $I_0 = 4.95 \times 10^{23}$ W/cm$^2$, \textcolor{black}{might soon be realizable~\cite{Yoon21}},  ($a_0=e E/m_e\omega c =600 $), where $e$ is the electronic mass, $\omega$ the laser frequency and $c$ the velocity of light in vacuum. The polarization of the laser is chosen to be $s-$polarized as \textcolor{black}{here our 2D simulations are then closer to 3D scenarios} as opposed to $p-$polarized laser light which can artificially heat electrons and can exaggerate the effectiveness of ion acceleration in such a scenario~\cite{Stark2017,Tamburini2012,Tamburini2010,Tamburini2011}. It interacts with a pre-formed fully-ionized Carbon plasma ($C^{6+}$) with a temperature $T_{e^-}=T_{C^{+}}=1200$ eV and density, $n_{e^-}=200n_c$, where $n_c = m_e \omega^2 / 4 \pi e^2$ is the classical critical density of a plasma for $1\,\mu$m laser wavelength. The target has a thickness of $0.6\, \mu$m, and is located at $12\, \mu$m from the left boundary of the simulation box. We employ transmitting and periodic boundary conditions in $x$ and $y$ direction, respectively. The simulation box has dimensions of $L_x \times L_y = (50\mu \textrm{m} \times 8\mu \textrm{m}$), with the cell size $\Delta_x \times \Delta_y = (10 \textrm{nm} \times 10 \textrm{nm}$) using $85$ particles per cell. \textcolor{black}{Laser-solid pair creation by QED processes mediated in Coulombic fields such as Bethe-Heitler~\cite{BetheHeitler.1934} and Trident processes~\cite{Bhabha.1935,Shearer} are not considered in these simulations. This should be reasonable as the ratio of the electric field strength of the laser to that of the atomic nucleus at ionic Debye length is $10^3$ (using average fields at a Bohr radius for Z=6 being $\langle E \rangle \sim 4\times10^{14}$V/m~\cite{BEIER200079}), favouring pair creation by photon-laser interaction over photon-nuclear interaction. Also with the sub-micron target of ion density of $(200/Z)n_c$, the pair creation probability due to the ionic nuclear field should be lower as also in Ref.\cite{He.2021}.}
We also performed parameter scans with the same laser but different target densities $[60, 100] n_c$ and observed a similar improvement by QED effects only for $100 n_c$. However, in a near-critical thin target $0.6\mu \rm{m}, a_0= 540, n_e=500n_c$ QED effects were observed to reduce ion energies.

\section{Dynamics}\label{SecEarlyStage}
 
The laser field pushes hot electrons inside the target forward that quickly reach the non-irradiated side (rear) of the target faster than the ions. This sets up a very brief TNSA field there which kickstarts the ion acceleration from the TNSA mechanism at around 50 fs. 
The electrons oscillate with relativistic velocity and thus, the effective critical density is reduced.
   \begin{figure}
    \centering
    \includegraphics[width=75mm]{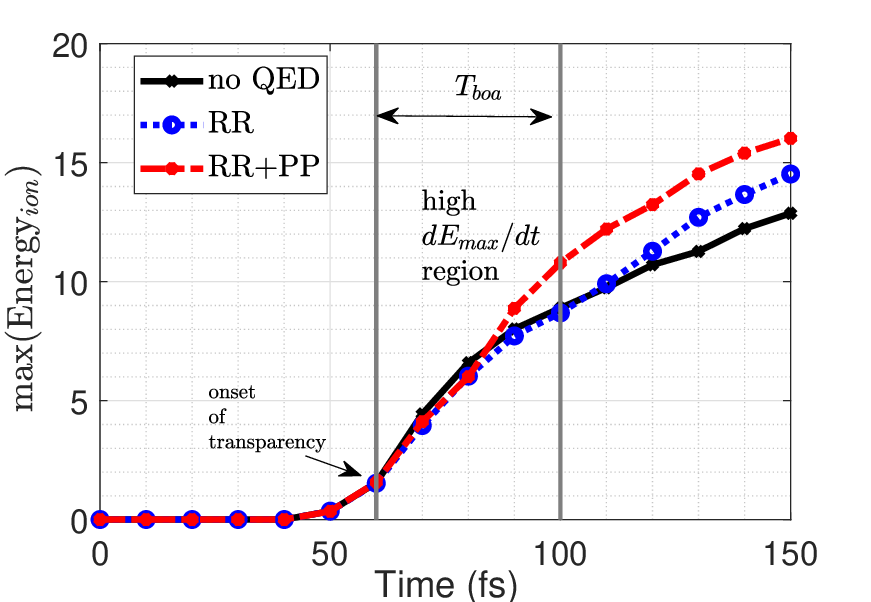}
    \caption{The 1D plot shows the maximum energy gained by ions $E_{max}$ with time $t$ in all three cases labelled. Here, a region in time is identified as $T_{boa}$ which starts at the onset of transparency and extends till the enhanced ion acceleration slows down (after which the slope of maximum ion-energy begins to change to a smaller value).}\label{rbi}
    \end{figure}
  \begin{figure}
    \centering
    \includegraphics[scale=0.3]{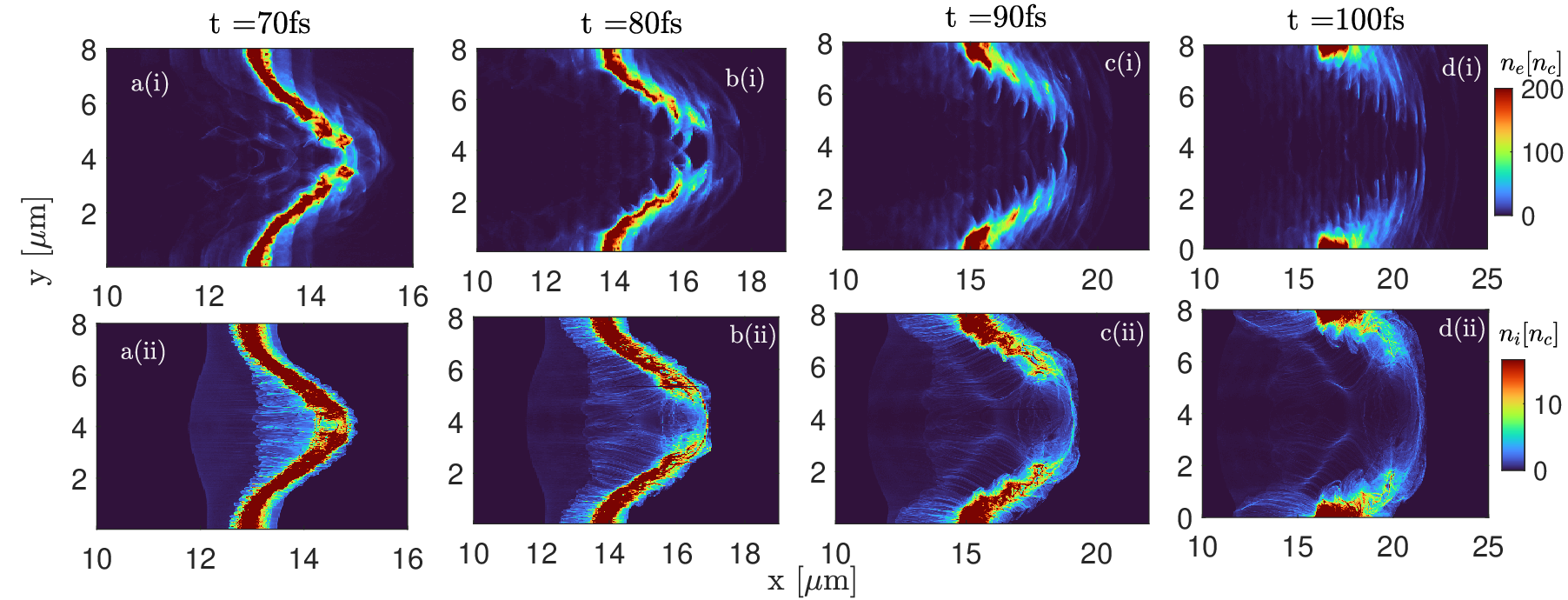}
    \caption{ These subplots shows 2D spatial distributions of electrons [top row, a-d(i)] and ions [bottom row, a-d(ii)] in the $T_{boa}$ region (only no-QED case shown).}\label{rbi2}
    \end{figure}
Moreover, as the recirculating hot electrons heat the target up, it begins to expand and the density lowers further. The target then begins to get relativistically transparent and the laser is able to penetrate through it. 
\textcolor{black}{This marks the onset of transparency (at $\sim60$ fs) where the streams of electrons and ions co-move with the penetrating laser and are susceptible to occurance of Breakout Afterburner.}
\textcolor{black}{In a realistic scenario of laser interacting with a thin foil, there can be multiple co-existing accelerating fields/mechanisms which can be broadly disentangled in time.}

\subsubsection*{Timing of the dynamics}
\textcolor{black}{In Fig.~\ref{rbi}, we plot the maximum energy gained by ions ($E_{max}$) as a function of time ($t$) in all three cases which are labelled \textcolor{black}{[when QED effects are artificially turned off, when only RR is included and when pair production is also included RR+PP]. The angular distribution of ions is not contained in this figure and one can not distinguish the on-axis and off-axis ions here. Yet, we broadly identify three stages of ion-acceleration.} Stage 1 is pre-transparency time (up till 60 fs) when the target is still intact and ion acceleration occurs with the combination of TNSA and RPA very briefly. After 60 fs, ion acceleration enters Stage 2 which we refer to as BOA-phase (marked as $T_{boa}$). Here electrons and ions co-stream with the laser and the system \textcolor{black}{could be} susceptible to RBI. This can also be seen in Fig.\ref{rbi2} where the 2D spatial distribution of electrons [$n_e(x,y)$ on top row, a-d(i)] and ions [$n_i(x,y)$ on the bottom row, a-d(ii)] in the $T_{boa}$ region is presented. \textcolor{black}{Here we only present the time evolution of electrons and ions in $(x$-$y)$ space for no-QED case to describe the timing of the dynamics. The QED effects were not very well distinguishable in this space. As will be seen in later sections the QED cases are clearly distinguishable from the no-QED case when visualised in angular-energy space.} \textcolor{black}{It is clear from these subplots in Fig.\ref{rbi2} that the electrons and ions are in close spatial proximity in Stage 2. The relative velocity between electron and ion flows acts as a source of free-energy for low-frequency electrostatic modes such as RBI to develop. In Stage 2 this growing mode is where ions \textcolor{black}{could} get accelerated from. Due to the Gaussian spatial profile of the laser, the electrons and ions stream at an angle as they move slightly away from the focal spot region.}
Stage 2 is characterised by a high rate of change in maximum ion energies and extends till the enhanced-ion-acceleration slows down (after around 100fs the slope of maximum ion-energy begins to clearly change to a smaller value). 
Stage 1 witnesses large production of high-energy-photons and pairs which saturates in Stage 2 (also see App. Fig.~\ref{time}). Afterwards, ion acceleration enters into Stage 3, where electrons get significantly expelled and acceleration occurs due to Coulomb explosion as also seen in ~\cite{Bulanov.2008}. In this paper, we focus on Stage 2 of ion acceleration as this is not only the stage of rapid energy gain dominating the overall accelerating mechanisms, but also the stage where QED effects reverse their energy-reduction-trend from its preceding stage. } 
\subsection{Early stage dynamics}\label{stage1}
 \subsubsection*{Electrons} 
\begin{figure}
\includegraphics[width=0.5\textwidth, height=.2\textheight]{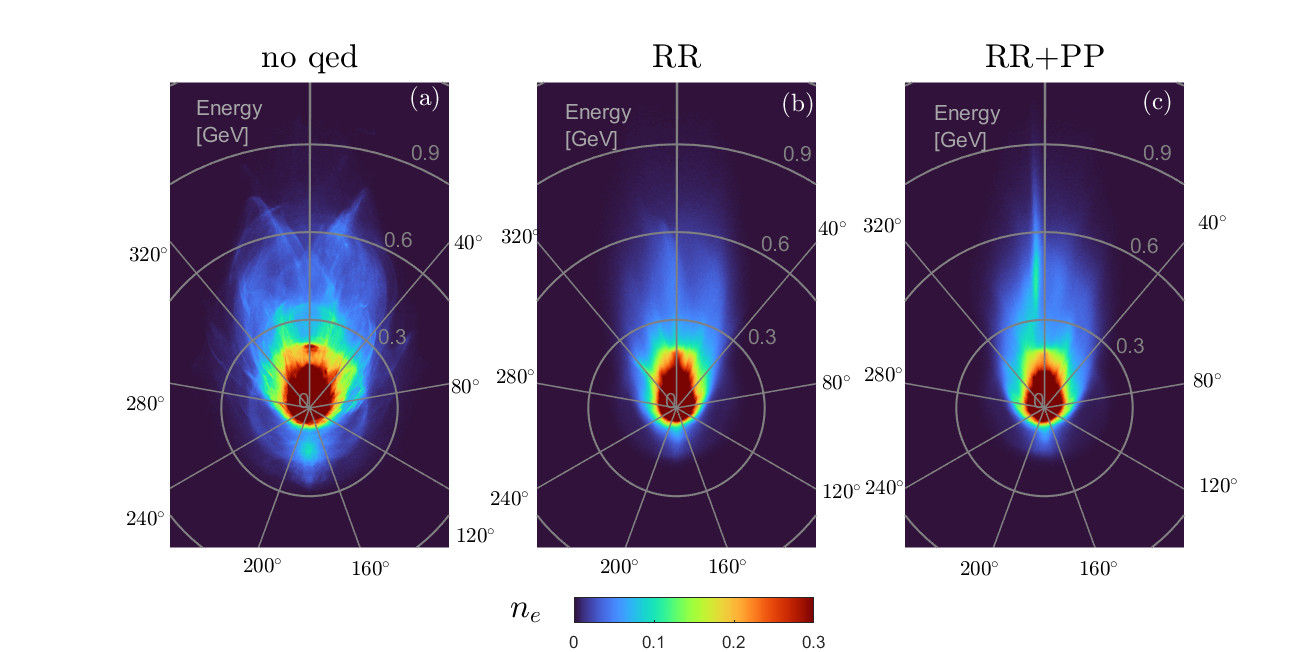}
\caption{Energy-angular distribution of electrons [in a.u.] in the BOA phase without radiation reaction [panel (a)], with radiation reaction [panel (b)] and with pair production as well [in panel (c), excluding the produced Breit-Wheeler electron density] at 80 fs.}
\label{fig-ne}  
\end{figure}
Fig.~\ref{fig-ne}, shows the electron's energy-angle distribution at 80 fs (BOA phase) where the laser pulse has already penetrated the target \textcolor{black}{(injecting electrons into vacuum laser acceleration by relativistic transparency~\cite{Singh2022})}. Panel (a) shows the case where the QED effects are artificially turned off, panel (b) shows the case when RR is included in the plasma dynamics and panel (c), when both RR and PP are included. One can clearly see in panel (a) that electrons stream diffusely at an angle and gain energy. The majority of the electrons stream in the forward direction (laser-propagation direction) and a small percentage of electrons also gain energy at the back ($\sim 180^\circ$). In panel (b), when RR is also included, the electrons become more forward-directed and the backward acceleration is suppressed. The latter observation is expected i.e. the electrons that counter-propagate the incoming laser experience Doppler-upshifted fields leading to a substantial suppression of its backward acceleration (also observed in Ref.~\cite{min.chen.iop.2010,Tamburini2010,Tamburini2011,Bhadoria2019}). As the laser-accelerated electrons lose part of their energies in high-energy-photon emission, the overall divergence of the electron's angular distribution reduces as they get pushed forward with the laser. Similar reduction in the electron's transverse momentum and electron cooling due to RR is also seen in Ref.~\cite{Tamburini2010,Tamburini2011,Gong2019, Capdessus2015}. Although laser collision with an electron-beam with quantum RR is shown to increase the electron energy distribution~\cite{Neitz.2013}, \textcolor{black}{here the overall impact is not dominated by stochasticity (See next section).}
\subsubsection*{Stochasticity in RR case}\label{App-Stochaticity}
\begin{figure}
\centering
\includegraphics[width=95mm]{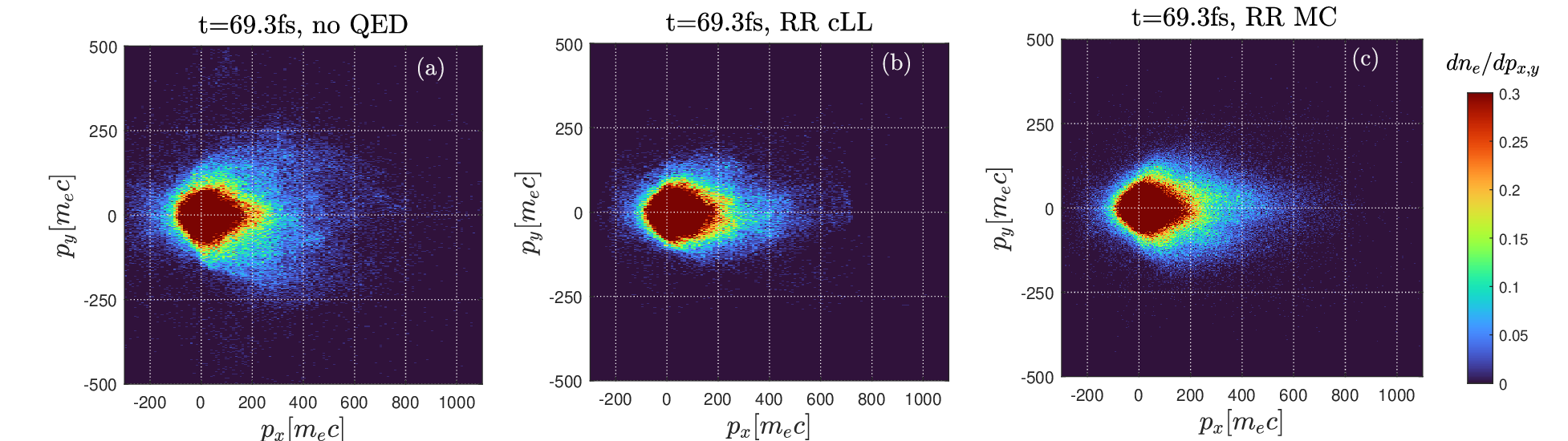}
\caption{The electron phase space in no QED case [panel (a)], RR modelled by corrected Landau-Lifschitz (LL) [panel (b)] and RR modelled by Monte-Carlo methods [panel (c)] at the onset of BOA phase.}\label{pxpy}
\end{figure}

 \textcolor{black}{In order to isolate the stochastic aspect of radiation reaction (RR) from only the continuous frictional drag on particles, we carried out one simulation which models RR with a corrected Landau-Lifschitz model that excludes the stochastic nature of photon emission (using Smilei code). Fig.~\ref{pxpy} shows the electron's momentum-phase-space distribution in the no-QED case [panel (a)], with RR modelled by corrected-Landau-Lifschitz [panel (b)] and RR modelled by Monte-Carlo methods [panel (c)].}

\textcolor{black}{ Comparing panels (b) and (c) of Fig.~\ref{pxpy} we see that a significant reduction in the electron's transverse momentum with RR is common in both. Panel (c), that also captures stochastic effects of RR seems to extend electron's momentum in both longitudinal as well as transverse direction. This is actually consistent with \cite{Neitz.2013} which shows that stochasticity leads to a greater spread of the electron energy distribution. Clearly in this scenario, the collimation of electrons due to the leading term of Landau-Lifschitz RR force (``drift term") dominates over the spreading out of electrons due to the stochastic (``diffusion term") effects, such that, compared to the no-QED case there is an overall collimation of the beam. \textcolor{black}{The subsequent ion energies due to stochastic effects is discussed in a later Appendix~\ref{Stoch_ions}}}


\subsubsection*{Additional pair plasma}
In Fig.\ref{fig-ne} panel (c), when RR+PP both are included, apart from a more collimated stream of electron fluid, here one can also see a higher density of electrons that also gain larger energy \textcolor{black}{(see around 0.6 GeV)}. This is due to the production of the BW-pairs that occurs due to the interaction of laser photons with the emitted gamma-ray photons.  \textcolor{black}{One can clearly see that the created pairs have higher maximum energy than target electrons.} The angularly streaming target-electrons gain more energy from the newly formed energetic pair-plasma at $0^\circ$ as all species of similar masses exchange energies. \textcolor{black}{This leads to additional collimation of the electron stream with the production of pairs.}

\begin{figure}
\includegraphics[width=0.5\textwidth, height=.2\textheight]{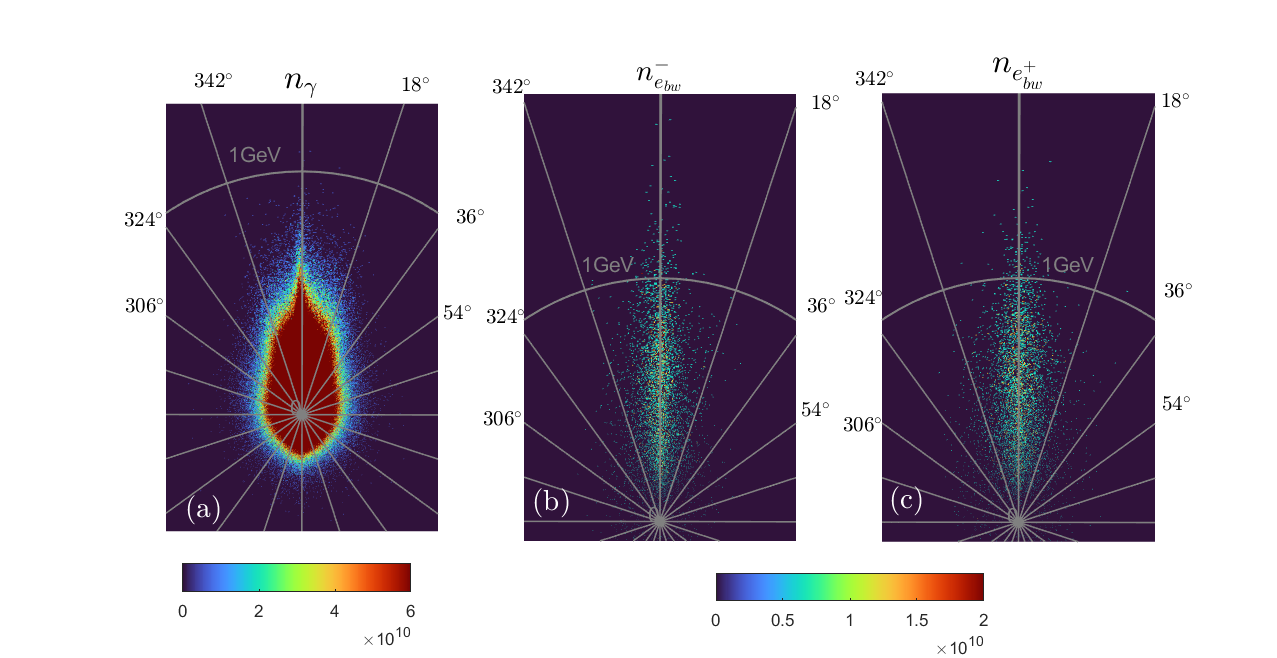}
\caption{Energy-angular distribution of photons [panel (a)], with BW electrons [panel (b)] and BW positrons [in panel (c)] at 80 fs.}
\label{fig-produced}  
\end{figure}

Fig.~\ref{fig-produced} (a, b and c) show the energy-angle distribution of photons, BW electrons and BW positrons respectively in the RR+PP case at 80 fs. One can clearly see a large number of gamma-ray photons in the laser-propagation direction being produced in panel (a) as the target turns transparent and the laser is allowed to interact with prolific electrons. In panel (b-c), we see the high-energy and forward-streaming pair-plasma that is responsible for the higher energy and density of electrons in Fig.~[\ref{fig-ne}(c)]. Since the target is already transparent, these pairs do not accumulate at the target region and are unable to shield the incoming laser as in the cushioning scenario \cite{Kirk_2013}, rather stream forward with the laser \textcolor{black}{pulse} and the ambient plasma. 
 \subsubsection*{Ions}
\begin{figure}
\includegraphics[width=0.5\textwidth, height=.18\textheight]{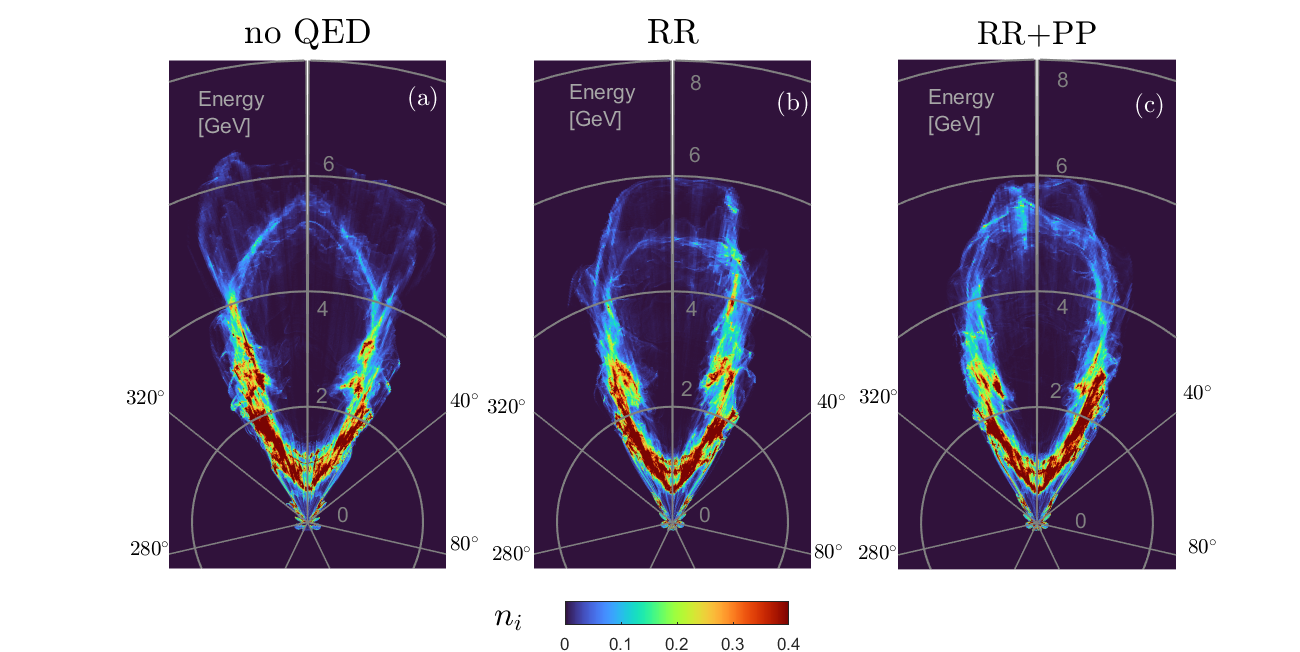}
\caption{Energy-angular distribution of carbon ions in the BOA phase without radiation reaction [panel (a)], with radiation reaction [panel (b)] and with pair production as well [in panel (c)] at 80 fs.}
\label{fig-ni}  
\end{figure}
Fig.~\ref{fig-ni} shows the ion distribution in the same fashion as in Fig.~\ref{fig-ne} and at the same time. In \textcolor{black}{the} no-QED case in panel (a) of Fig.~\ref{fig-ni}, the ions with the highest energy (around $ 6.5$ GeV) are off the axis of laser polarization or propagation, as also seen in Ref.\cite{Yin2011}. This occurs at 80 fs when RBI \textcolor{black}{could} operate which is a low-frequency high-amplitude electrostatic mode that feeds on the relative flow velocity between electrons and ions and accelerates ions with a wave-particle resonance mechanism~\cite{Albright2007,Stark2018}. In the same figure, one may also see some ions with $\sim 5.7$ GeV energy which are on the laser-propagation axis. This is when the off-axis ion streams mutually interact. In Fig.[\ref{fig-ni}(b)], the highest gain in energy and the angular divergence of these high-energy ions is reduced at 80 fs when BOA mechanism \textcolor{black}{could be} at play. The on-axis and off-axis ions gain nearly the same energies in this case. Further, in Fig.[\ref{fig-ni}(c)], the angular divergence of the ions is even smaller, and the on-axis ions gain much higher energy ($\sim 5.8$ GeV) than the off-axis ones ($\sim 4.6$ GeV). The high-energy, on-axis ion bunch is accelerated due to a similar electron bunch in  Fig.[\ref{fig-ne}(c)] on account of the pair plasma Fig.[\ref{fig-produced}(c)]. The role of an RBI in these bunches of high-energy ions \textcolor{black}{seems relevant} in higher acceleration of ions. 
The expanding TNSA ions, target electrons and the BW pairs stream forward with the laser and the free energy in the particle streams gives rise to electrostatic mode, RBI, that resonates with ions allowing them to be rapidly accelerated. The energy loss by electrons is constantly filled up by the long-pulse laser.
This beam-like expanding plasma is susceptible to the growth of RBI where the phase velocity of the instability is comparable to the highest accelerated velocities of the ions~\cite{Albright2007}.

\textcolor{black}{It should be noted that this scenario \textcolor{black}{could be similar to }that of Directed Coulomb Explosion~\cite{Bulanov.2008} where RPA precedes the later Coulomb explosion stage for acceleration of ions.} \textcolor{black}{Though here a higher transparency with higher $a_0$ with transverse target expansion would reduce RPA's efficiency~\cite{Bulanov.2016}, there may be a more complex interaction here with a phase of hybrid-RPA-BOA accelerating ions from the off-focal opaque part and the focal transparent part of the target, respectively. Confirming an exact composite-accelerating-mechanism calls for an investigation of shorter time-scale particle dynamics for classical case itself, especially with QED effects enhancing ion energies. However, here we limit ourselves only to a discussion on the analysis of longitudinal electrostatic field structure (similar to Ref.~\cite{Yin2007,Albright2007,King2017}) where we look for existence of signatures of RBI.}
\begin{figure}
\adjincludegraphics[height=0.14\textheight,width=0.53\textwidth,trim={2.75cm 0 0 0},clip]{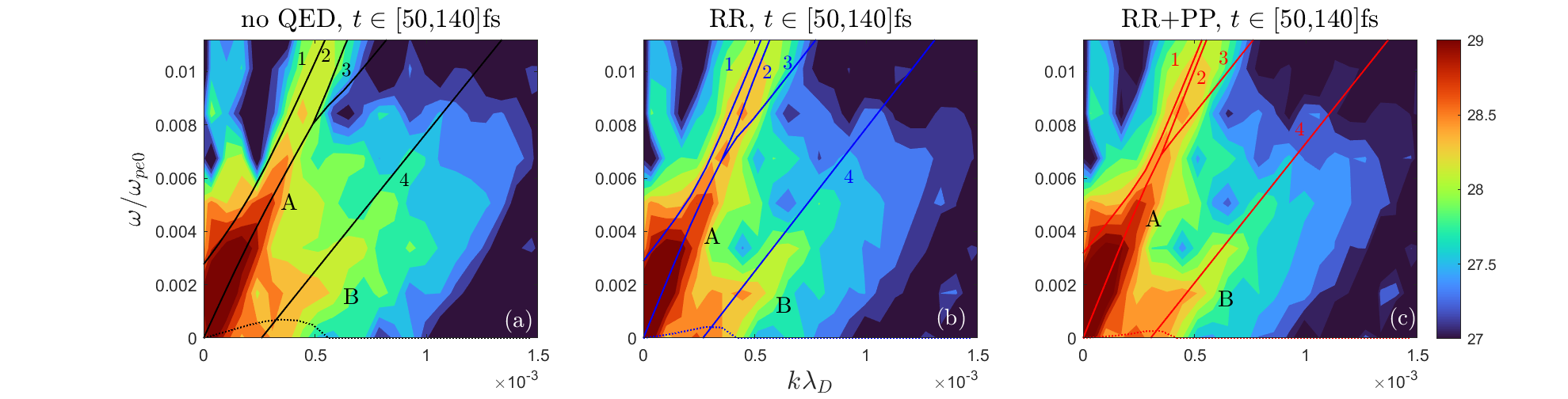}
\caption{Spectral power as a function of wave number (normalized by Debye's length \textcolor{black}{with initial temperature}) and frequency (normalized by plasma frequency), $ |E_x(\omega,k)|^2$, in log scale for $t\in [50-140]$ fs and $x \in [10-50]\, \mu$m for all the three cases [no QED panel (a), RR panel (b) and RR+PP panel (c)] obtained from the simulations. The real and imaginary roots of Eq.\ref{eqn-dr} (solid and dotted respectively) are over-plotted to facilitate comparison. }
\label{fig-fft}  
\end{figure}
\section{Transparency stage}
\subsubsection*{Identifying RBI from simulation}
The Fourier analysis of the longitudinal electric field from the simulations can shed light on the electrostatic structure of the accelerating fields in the transparency region. This has been performed for all three cases and is shown in Fig.~\ref{fig-fft}. Panel (a),(b) and (c) show $ |E_x(\omega,k)|^2$ in log scale for the no QED case, with RR and with RR+PP respectively. The Fourier window has been chosen to be $[50-140]$ fs and $[10-50]\, \mu$m to capture the salient features of the instability dynamics in the BOA phase. The BOA time window ($t_{\rm BOA}\in [60-100]$ fs) is identified by the time when we observe rapid ion acceleration ($\sim 2-4$ times in every 10 fs) in our simulations, after which the rate of ion acceleration becomes smaller ($\sim 1-1.1$ times in every 10 fs, \textcolor{black}{as seen in Fig.~\ref{rbi}}). This BOA window is well within the resolution of the Fourier window shown in Fig.~\ref{fig-fft}. In this power spectrum in Fig.~\ref{fig-fft}, two distinct low frequency branches can be clearly identified in all three panels [(a)-(c)]. Clearly, one primary branch (labelled $A$) has a higher slope and energy than the other (labelled $B$). The primary branch $A$ 
intersecting the origin \textcolor{black}{is identified as the} the growing RBI~\cite{Yin2007}. This branch could also be clearly identified even when we chose smaller windows at earlier times like $t \in [50-80]$ fs or $t \in [50-100]$ fs (not shown here), with lower phase velocities than the ones shown here. The phase velocity of this branch is seen to increase as we increase the temporal fourier window within $t_{BOA}$, consistent with Ref.~\cite{Stark2018}. The lower, diffuse and less-powerful branch $B$ appears only some time after ($t \in [50-90]$ fs onwards) the appearance of the primary branch. These two branches merge slightly in panel (a). Looking at panels (b) and (c), one can broadly see that the branch $A$ is more powerful in both QED cases than in panel (a) [even more with RR+PP case]. Moreover, the branch B becomes notably more distinct in panel (b) and marginally even more in panel (c). This may be due to the fields generated by the angularly drifting plasma streams that mutually interact leading to the high-energy on-axis ions seen in the tip of a bubble-like form that ions make in Fig.~\ref{fig-ni} (potentially a mode harnessing the free energy in off-axis high energy ion streams). As the radiatively cooled electrons become more forward-directed in QED cases (Fig.~\ref{fig-ne} [(b-c)]), the angular separation between the streaming plasma ions also lowers. This allows more interaction between the streams and thus the branch $B$ becomes more distinct. An additional lowering of this angle due to pairs produced at $0^\circ$ makes this branch $B$ stronger in Fig.~[\ref{fig-fft}(c)].
 \subsubsection*{RBI from linear theory}
 The dispersion relation of RBI\cite{Albright2007} from the linear kinetic theory assuming cold angularly streaming plasma for the instability is given as 
\begin{equation}\label{eqn-dr}
\sum_{s=e,i} \frac{\omega_{p,s}^2[1+(p_{s}\sin\theta_s/m_sc)^2]}{k^2\gamma_s^3(v_{\phi} - v_{s}\cos\theta_s)^2}  = 1,
\end{equation}

where, $\omega_{p,s}$ is plasma frequency, $v_s/p_s$ are the stream velocity/momentum,$v_\phi$ is the phase velocity ($\omega/k$), $\gamma_s$ the respective Lorentz factor and $\theta_s$ is the angle of drift, with $s=e,i$ denoting the electronic and ionic streams respectively. \textcolor{black}{Although one can not deny that perturbative approach might not be the most sophisticated approach to study this but it is the best non-simulation approach available that can facilitate a deeper understanding of such a complex interaction.} 
The dispersion relation of this instability in Eq.~\ref{eqn-dr} has been solved and the 4 roots of the quartic equation have been overplotted in Fig.~\ref{fig-fft}. The input plasma parameters (average electron and ion density, angles of streams and energies) have been extracted from the simulations in each case at around 70 fs when BOA \textcolor{black}{could be} active (see App.~\ref{App-plasmaChar})~\footnote{$[n_{e},\theta_e,\epsilon_e]^0$=$[0.0026n_0,24^\circ,0.3420$GeV], $[n_{i},\theta_i,\epsilon_i]^0$=[0.0404$n_0,16.9^\circ,1.914$GeV]; $[n_{e},\theta_e,\epsilon_e]^{RR}$=[0.0046$n_0,8.2^\circ,0.3925$GeV], $[n_{i},\theta_i,\epsilon_i]^{RR}$=[0.049$n_0,13.8^\circ,2.03$GeV];
$[n_{e},\theta_e,\epsilon_e]^{RR+PP}$=[0.0032$n_0,5.1^\circ,0.4128$GeV], $[n_{i},\theta_i,\epsilon_i]^{RR+PP}$=[0.0591$n_0,17.9^\circ,2.01$GeV],

$n_{(e,i,0)},\epsilon_{e,i}$ being the electron, ion and the initialised particle density and energies respectively}. It should be noted that we use the same dispersion relation for QED cases (Fig.~\ref{fig-fft} [(b-c)]) as well. This is reasonable as we carefully choose the plasma parameters at the time after the production of photons and pairs has mostly saturated \textcolor{black}{see also Sec.~\ref{SecEarlyStage}}. \textcolor{black}{Major} impact of RR and pair plasma are still well captured in the form of changes in the plasma distribution function extracted from the simulation that already includes probabilistic photon emission in plasma evolution.

There are 2 real and 2 complex roots of this equation. One high frequency real root (starts with positive-frequency as also in Ref.~\cite{Albright2007,Stark2018a}) and the other low frequency real root (negative frequency at $k=0$ crosses the $\omega=0$ axis as the wave-number increases). The other 2 roots are complex conjugates with the same $\Re(\omega)$ till the non-zero imaginary part vanishes, after which the real parts bifurcate. The positive imaginary part (dotted line in Fig.~\ref{fig-fft}) is the unstable mode while the negative (damped mode) is not shown here. 
A good match between the branch $A$ (from simulation) and the real part of growing complex root from linear kinetic theory (overplotted solid line) is visible in all 3 panels. The phase velocity of the primary branch ($v_p\sim0.84c$ in Fig.~\ref{fig-fft}(a)) is comparable to the ion velocities attained by the off-axis ions ($v_i=0.86c,\epsilon_{i}\sim11.26$ GeV) \textcolor{black}{hinting to the possibility
} of the instability \textcolor{black}{playing a role }in the ion acceleration. The instability growth rate progressively lowers and the bifurcation of the roots shifts to lower $k$ values respectively, which is as expected \cite{Stark2018,Stark2018a}. A lower angle and higher energies of the electron stream (see Fig.~\ref{fig-ne}) is also shown to enhance the phase velocities of the RBI wave~\cite{Stark2018}. \textcolor{black}{As we see already in Sec.~\ref{SecEarlyStage} that RR and RR+PP lead to a much more collimated plasma stream in stage 1 of acceleration (Sec.~\ref{SecEarlyStage}), a higher phase velocity of RBI with these QED effects is understandable.}
Thus, from the spectral plots it is clear that RR and RR+PP \textcolor{black}{would} enhance the RBI on account of a radiatively cooled more-forward-directed electron and ion beam. Interestingly, the low-frequency real root of the same dispersion relation, which has negative frequency for $k=0$, matches very well the branch $B$ picked up by the FFT of the longitudinal electric fields from simulation. 
This points to a lower ion-mode that additionally bestow the high-energy on-axis ions, accelerated at later-time due to mutually interacting angular ion streams. \textcolor{black}{In the dispersion relation of RBI, with angularly streaming plasma characteristics extracted from 70 fs, when the electronic contributions are allowed to vanish, $n_e=0$ and $\epsilon_e=0$, we obtain a quadratic equation giving 2 real roots. One of the real roots ($\omega_r$) of the perturbation, which would mean non-growing/non-damping oscillation, matches perfectly with the lower frequency root $4$ in Fig.~\ref{fig-fft} that lies over the branch B. This branch gets stronger with QED effects, which hints to growing oscillations between ion-streams as they become more forward directed. These ion oscillations bring the outward bursting ions more towards the axis of laser propagation.}
\begin{figure}
\includegraphics[width=0.5\textwidth, height=.2\textheight]{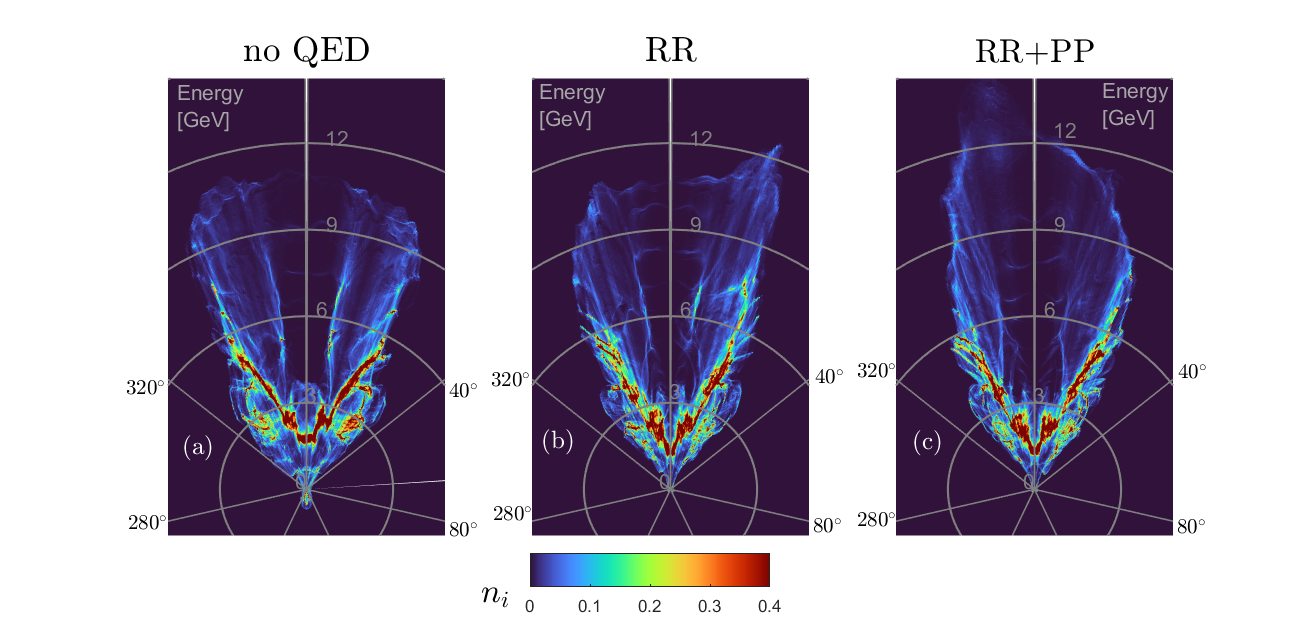}
\includegraphics[width=0.4\textwidth, height=.1\textheight]{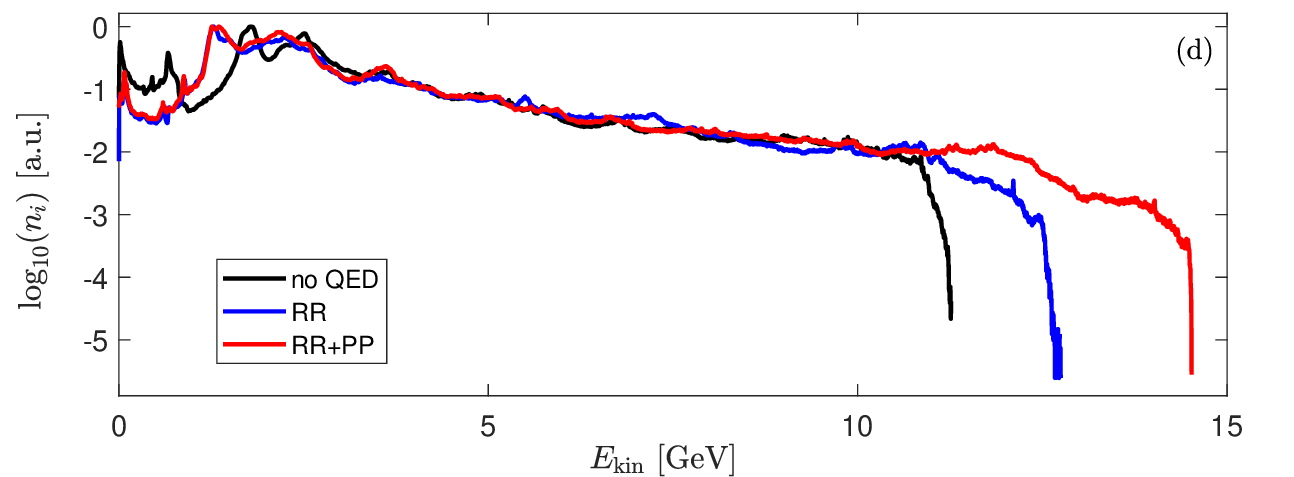}
\caption{Energy-angular distribution of carbon ions in the BOA phase without radiation reaction [panel (a)], with radiation reaction [panel (b)] and with pair production as well [in panel (c)] at 130 fs. Panel (d) shows angle-averaged ion energy distribution at the same time.}
\label{fig-ni-late}  
\end{figure}
\subsubsection*{RBI phase velocity and resonant Ion velocities}
Fig.~\ref{fig-ni-late} shows the angle-energy and the $\theta$-averaged ion-distribution in the three cases at a later time of 130 fs. The highest energy gained by the ions at this time in the no QED case is $\sim 11.26$ GeV, with RR $12.7$ GeV ($\sim 12\%$ higher) and with RR+PP it is 14.52 GeV ($\sim 30\%$ higher). Corresponding ion velocities $v_i=[0.86c,0.88c, 0.90c$] are in good agreement with the respective phase velocities of the RBI $v_p\sim[0.84c,0.89c,0.91c]$ from simulations [branch A] and also with the overplotted phase velocities of the RBI from linear theory $v_p\sim[0.75c,0.82c,0.90c]$. This \textcolor{black}{presents some hints on possible}
wave-particle acceleration mechanism~\cite{Albright2007}(see more details in App.\ref{App-PhaseVelocity}). \textcolor{black}{A good agreement can be seen even without including the RR term in the instability calculation because the strong impact of RR in Stage 1 is actually included via simulations in the form of changes in the distribution function in Stage 2 of the instability development.}
\section{Conclusion}
In conclusion, we investigated the effect of radiation reaction and pair production on the ion acceleration where the BOA mechanism \textcolor{black}{may }operate. We demonstrate how QED effects can impact the collective plasma behaviour in the early stages of laser-plasma interaction. \textcolor{black}{This may }lead to an enhanced phase velocity of RBI in a later BOA stage. \textcolor{black}{Though, the spectra presented here could also be explained by RPA mechanism by taking into account transverse expansion of the target~\footnote{Private communication with S.S. Bulanov and S.V. Bulanov}, and a more systematic study of ion electron and ion phase space at smaller time scale to search for signatures of Relativistic Buneman Instability could further clarify the nature of accelerating mechanism.}
\textcolor{black}{Nonetheless, non-classical effects clearly modify the plasma distribution significantly in this regime and can lead to a gain of higher energy (around $30\%$) by the ions.} The angle of streaming between the transparent target electrons and the forward-directed $e^-e^+$ pair-plasma plays a principal role in plasma dynamics and the consequent high ion energy gain. \textcolor{black}{Measuring the deviations in the experimentally observed particle spectra from classically expected results one can help identify or verify QED signatures. Apart from energy enhancement, simulations presented here also show that with QED effects the highest ion-energy signal would be for the particles directed near the laser propagation whereas without QED this would be at an angle appreciably above zero. This can be a key signature to verify QED effects.} Recent related experimental corroboration of QED effects~\cite{Poder2018,Sarri2015,Cole2018} and the advent of ultra-high intensity lasers~\cite{Yoon21,papadopoulos2016,Gales2018,danson2019} places these findings in very exciting times. 

\subsection{Acknowledgements}
The authors would like to thank Naveen Kumar for suggesting to look into Breakout Afterburner mechanism for ion acceleration and for his follow up insightful discussions during the early stage of this work. The authors would like to thank Tom G. Blackburn, Arkady Gonoskov, Joel Magnusson, Brian Reville and Matteo Tamburini for their insightful discussions throughout this work. 
\textcolor{black}{In particular, we are grateful to Stepan Bulanov and Sergey Bulanov for their insightful comments on the role of the Buneman instability and the influence of the RPA mechanism.}

\section{Appendices}


\subsection{Plasma characteristics extraction for instability calculation}\label{App-plasmaChar}

\begin{figure}
\centering
\includegraphics[width=65mm]{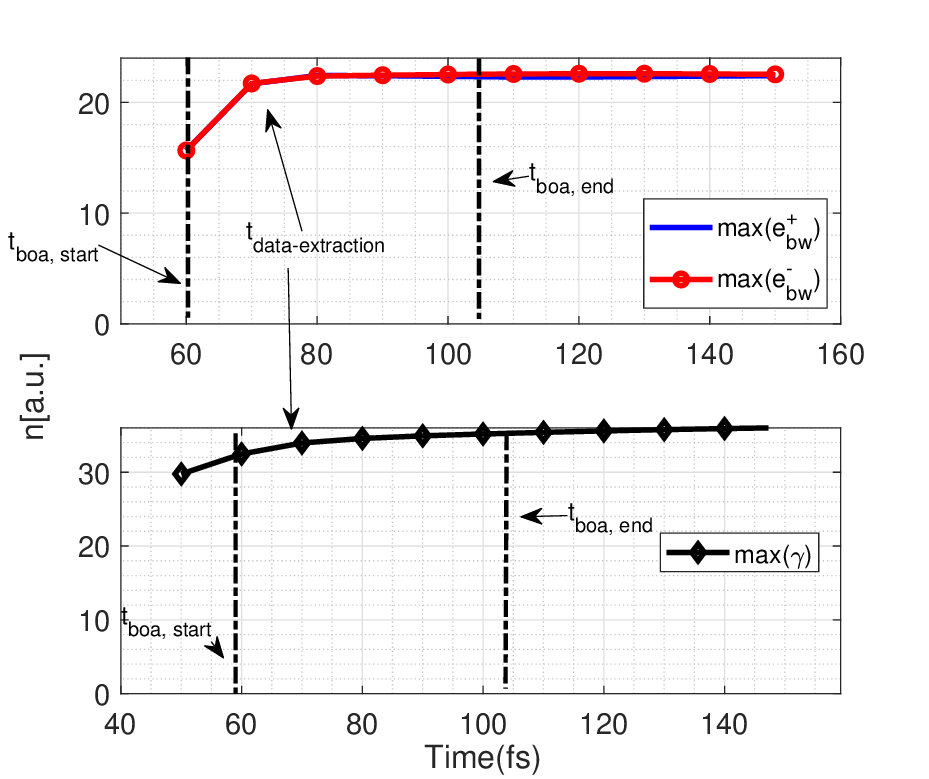}
\caption{Photons and pairs saturate after which the direct impact of QED effects can be assumed less significant.This justifies the dropping of RR term in Lorentz force from the Vlasov equation. QED effects still captured in form of changes in plasma distribution.} 
\label{time}
\end{figure}

To extract the plasma characteristics, we choose a time of 70fs. This is when the onset of RBI is expected with the target turning transparent and the electrons and ions streaming forward. Fig.~\ref{time} shows the time evolution of the peak values of the number of photons and pairs produced in the QED cases of the simulation considered in the manuscript. A clear saturation of the number of photons and pairs around 70 fs \textcolor{black}{(Stage 2)} implies that the emission of particles is negligible beyond this time and an RR term in the Lorentz force can be safely dropped from the Vlasov equation. It is after this time that the ions experience a boosted acceleration \textcolor{black}{potentially} due to RBI. The QED effects are strong at earlier times \textcolor{black}{(Stage 1)} and modify the plasma distribution function that would be used at 70 fs as the initial condition to the instability evolution.

\begin{figure}
\centering
\includegraphics[width=80mm]{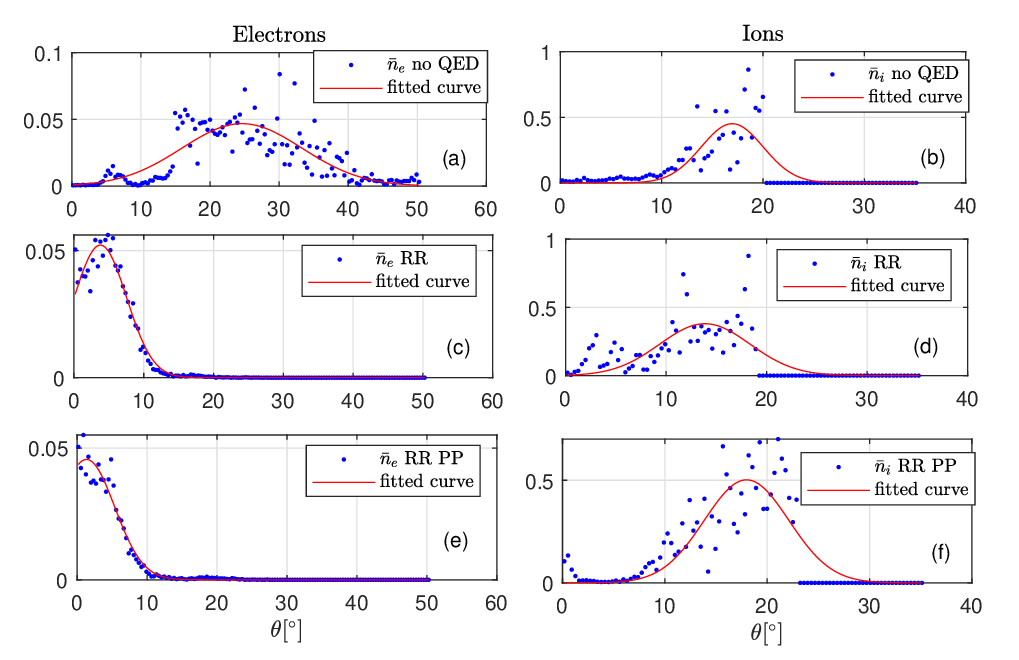}
\caption{Average number density of electrons (first column) and ions (second column) as a function of angle. }\label{fitting}
\end{figure}

To compute the dispersion relation for the RBI, an initial streaming plasma distribution function
\begin{equation}
    f_{e,i}(\vec p) = n_0\delta(\vec p - \vec{P_0}),
\end{equation}
with mean drift $\vec{P_0}=[P_{0x},P_{0y},0]$ (as in Ref.~\cite{Albright2007,King2017}), is perturbed with a small perturbation of the form $\xi=\xi_0 \exp \iota (\vec k.\vec x-\omega t)$. As PIC simulations generate a continuous distribution of plasma particles, to attain plasma values that can fit into this cold distribution we average out the distribution. 
For this, a normal distribution curve is fitted on to the particle energy distribution at each angle $\theta_i$ using the method of non-linear least squares that iteratively minimises the residue between PIC data and the fitted curve (with a goodness of fit of $R^2=[0.8-0.9]$), and the mean is extracted. With this, an average energy  $\bar E_{i}$ is determined with an uncertainty of $\pm \Delta E_i$ (limits of 95\% confidence interval). This energy and the corresponding particle number $\bar n_{i}$ with uncertainty $\pm \Delta n_i$ is then plotted as a function of $\theta$ and the mean angle of flow is determined [see Fig.~\ref{fitting}]. Since, the fast moving particles participate in the RBI, a cutoff of 0.1 GeV and 1GeV is applied on the electron and ion distribution respectively while fitting, to rule out the target species \textcolor{black}{far from the focal area} that are still opaque to the laser .

In this procedure of determining mean plasma density, energy and angle, the pairs population is also added to the electrons in the QED case. Positrons can also be added here due to presence of charge in the form of $e^2$ in the dispersion relation, Eq.~\ref{eqn-dr}. This addition does not manifest itself as a significant change in the number density of plasma, rather the average angle of flow of the electron cloud, making it more and more forward directed, as shown in Sec.~\ref{SecEarlyStage}.


\subsection{Numerical Robustness}\label{App-NumericalRobustness}
\begin{table}[]
\begin{tabular}{|l|l|l|l|} \hline
$N_{ppc}$ & $E_{max}$, no QED  & $E_{max}$, RR & $E_{max}$, RR+PP \\ \hline
20 & 11.05 & 13.1 ($\uparrow 18\%$) & 14.33($\uparrow 30\%$)\\ \hline
85$^{\rm{as\ in\  manuscript}}$ & 11.26 & 12.7($\uparrow 13\%$)  & 14.52($\uparrow 29\%$)\\ \hline
200 & 11.50 & 11.76 ($\uparrow 2\%$)  & 13.4($\uparrow 16\%$)\\ \hline
300 & 11.55 & 12.36($\uparrow 7\%$)  & 13.15 ($\uparrow 15\%$)\\ \hline

\end{tabular}
\caption{Maximum ion energies in GeV. The percentage change from the no-QED case is tabulated in round brackets in each QED case. All simulations are performed by Epoch.}\label{table1} 
\end{table}
\subsubsection*{Particle numbers}\label{nppc}
The 2D simulations were repeated with different numbers of macro particles per cell ($N_{ppc}$) to check numerical reliability of these results. The maximum ion energies are tabulated in Table.~\ref{table1}.The maximum energy gained by the ions varies with different numbers of macro-particles, $N_{ppc}$, even for no-QED case. With this we prescribe a numerical error-bar to the \textcolor{black}{value of maximum energy gained by the} ions. \textcolor{black}{At times there is an overlap between the lower-end of energy error-bar of QED case with the upper-end of energy error-bar of no-QED case (also including different $N_{ppc}$ values in between the ones in the table).} This can also be seen in table~\ref{table1} where the ion energy of 11.76 GeV in the RR case almost overlaps with the no-QED case of 11.55 GeV with a different number of particles.

\begin{figure}
\centering
\includegraphics[width=80mm]{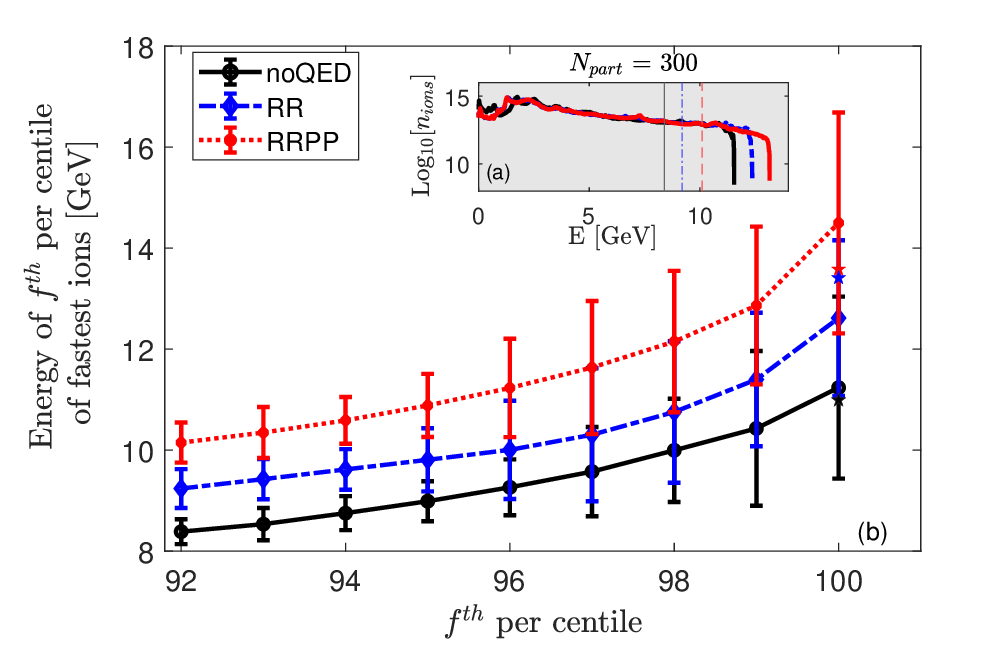}
\caption{Minimum energy of the fastest $1-8\%$ of particles of the high-energy tail with an errorbar due to different $N_{ppc}$.}\label{Percentile}
\end{figure}
To clearly disentangle these error bars, instead of comparing just the maximum ion energy gained by fastest ions, \textcolor{black}{we compare energy cutoff of} some $f^{th}$ percentile of the ion's high energy tail. \textcolor{black}{This can be seen in Fig.~\ref{Percentile}. Here the circular data point represents the case discussed in the paper with the error bar due to different numbers of particles. So, for instance, for $f=92$ the circular data point represents the energy cutoff between the fastest 8$\%$ of the particles and the remaining $92\%$ with lower energy. The corresponding error-bar originates from the same energy cutoffs of the fastest $8\%$ from the simulations with different numbers of particles. As expected, the energy cutoff of the fastest $8\%$ is lower (around 8 GeV) than the fastest $1\%$ (around 11 GeV). The top inset (a) of Fig.~\ref{Percentile} shows one typical ion energy spectrum with vertical lines marking this energy cutoff of the top $8\%$ ions from the high energy tail (instead of only the highest energy which shows much larger numerical variations due to the chosen number of particles).} The overlapping error-bar due to different particle numbers disentangles at around $92^{nd}$ percentile of particles
in high-energy tail, clearly corroborating an energy enhancement with QED effects by 2D simulations.

\subsubsection*{From an alternative code: Smilei}
\begin{figure}
\centering
\includegraphics[width=80mm]{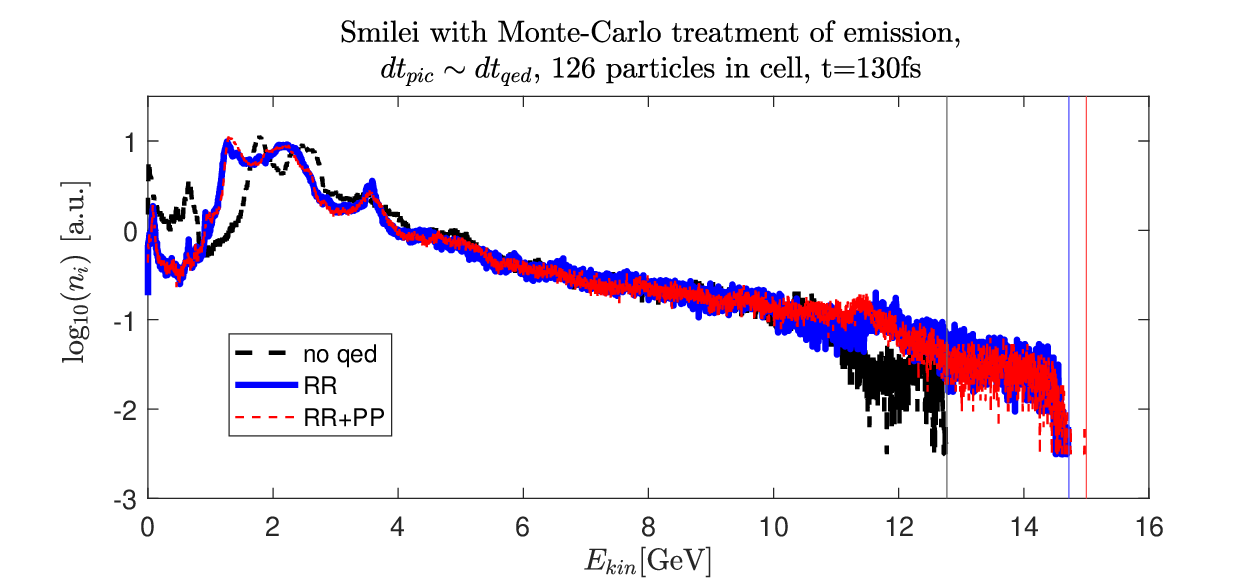}
\caption{Ion energy distribution at 130 fs from the exact same scenario in the manuscript when simulated by another PIC code Smilei~\cite{DEROUILLAT2018351} shows the same trend and is within the error bar of Fig~\ref{Percentile}}\label{smilei}
\end{figure}

Moreover, to cross check these findings, we also performed \textcolor{black}{few} additional simulations with another PIC code called Smilei~\cite{DEROUILLAT2018351} in light of some differences observed in the collisions modules of PIC~\cite{Bhadoria.Arxiv}. First one simulation is carried out with the same number of particles ($N_{ppc}=$84) and a spatio-temporal step as in the manuscript ($dt\sim0.012$ fs ). The peak energies are tabulated in the top row of Table~\ref{table_smilei} and show the same energy enhancement through QED effects.
Other simulations with higher numbers of particles and a smaller time step $dt_{pic}\sim dt_{qed}\sim0.007$ fs~\cite{Ridgers.2014} were also carried out and are tabulated further in Table~\ref{table_smilei}. A typical ion energy spectrum from the simulation with a faster time step and higher particle numbers (126) is shown in Fig~\ref{smilei}.  Here as well we see the same behaviour of energy enhancement with QED effects which are well within the error bar of Fig.~\ref{Percentile}.
\begin{table}[]
\begin{tabular}{|l|l|l|l|} \hline
$N_{ppc}$ & $E_{max}$, no QED  & $E_{max}$, RR & $E_{max}$, RR+PP \\ \hline
84$^{dt>dt_{qed}}$ & 12.40 & 12.61($\uparrow 2\%$)  & 13.87 ($\uparrow 12\%$)\\ \hline
126$^{dt\sim dt_{qed}}$ & 12.76 & 14.71($\uparrow 15\%$)  & 15.00 ($\uparrow 18\%$)\\ \hline
210$^{dt\sim dt_{qed}}$& 12.33 & 12.48($\uparrow 2\%$) &12.82 ($\uparrow 5\%$)\\ \hline
252$^{dt\sim dt_{qed}}$ & 11.23 & 12.55($\uparrow 11\%$)  & 12.71 ($\uparrow 14\%$)\\ \hline

\end{tabular}
\caption{Maximum ion energies in GeV. The percentage change from no QED case is tabulated in round brackets in each QED case. All simulations are performed by Smilei.}\label{table_smilei}
\end{table}
\begin{figure}
\centering
\includegraphics[width=80mm]{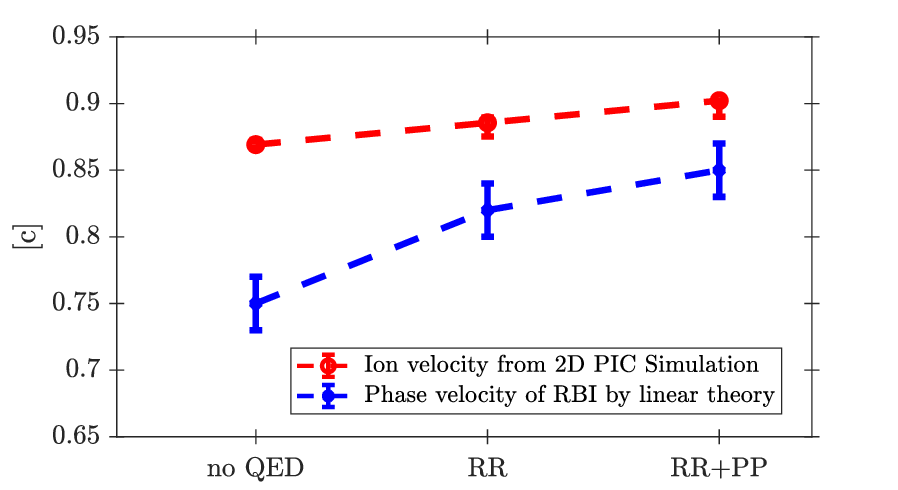}
\caption{Maximum ion velocities from 2D PIC simulations (red), with the error bar from different number of quasi-particles per cell. Phase velocities from RBI (blue), with error bars due to uncertainty in data extraction from simulation. It should be noted that the extraction of the plasma characteristics for instability's phase velocity calculation (blue error bar) has been obtained from the one simulation presented in the main text.}\label{errorbars}
\end{figure}
\subsubsection*{Spatial and Temporal resolution}
In the simulations in Sec.\ref{nppc} and that in the manuscript, the time step is chosen as $dt\sim0.012$ fs. This is larger than yet close to the photon emission time $dt_{QED}=0.007$ fs. This is reasonable because around this ratio of $dt/dt_{QED}\sim0.58$ it has been shown~\cite{Ridgers.2014} that the error in energy radiated as photons per particle in Monte Carlo simulations converges to a reasonable accuracy.  Nevertheless, we performed one simulation with $N_{ppc}=85$ [as in Sec~\ref{simulations}] but a much smaller time step $dt\sim0.006$ fs as well (Table.~\ref{table2}). The spatial resolution chosen as default by the Epoch code to ensure fulfilment of CFL criterion gives a cell size of 3.2 nm in both dimensions.  The peak energies from this are marked as a star in respective color in Fig.\ref{Percentile} (see on 100$^{\rm th}$ percentile line) and show around 23$\%$ energy enhancement. These values are still well within the prescribed error bar showing that a lower time step would not generate a complete outlier for a typical value of $N_{ppc}$. Also, in Sec.\ref{nppc} and the paper, the spatial resolution was held at 10 nm which resolves the electron skin depth of 11.25 nm. Additional sets of simulations with finer spatial resolution were also performed from both Epoch and Smilei codes. 
The ion energies show the same trend of energy enhancement and are tabulated in Table\ref{table2}.
\begin{table}[]
\begin{tabular}{|l|l|l|l|l|} \hline
$N_{ppc}$ & $dt,dx$ &$E_{max}^0$  & $E_{max}$, RR & $E_{max}$, RR+PP \\ \hline
85$^{\rm{Epoch}}$ & 0.007fs, 5.0nm & 11.93 & 12.63 ($\uparrow 5\%$) & 12.68 ($\uparrow 7\%$)\\ \hline
85$^{\rm{Epoch}}$ & 0.006fs, 3.2nm & 10.99 & 13.41 ($\uparrow 22\%$) & 13.58 ($\uparrow 24\%$)\\ \hline
126$^{\rm{Smilei}}$ &0.008fs, 6.0nm & 11.15 & 11.46($\uparrow 3\%$)  & 12.42($\uparrow 12\%$)\\ \hline
126$^{\rm{Smilei}}$ &0.008fs, 10nm& 12.76 & 14.71($\uparrow 15\%$)  & 15.00 ($\uparrow 18\%$)\\ \hline
\end{tabular}
\caption{Maximum ion energies in GeV with enhanced spatial resolution in Smilei code and in Epoch corroborate the trend of energy improvement. \textcolor{black}{Superscript `0' denotes the `no-QED' case here.} }\label{table2}
\end{table}


Thus, these 2D simulations successfully capture a clear 
trend of enhancement of ion-energy by QED effects.


\subsubsection*{Outlook}
\textcolor{black}{It should be noted that this work is at the front line of what can be implemented numerically at those extreme parameters including pair production and radiation reaction with corresponding high numerical uncertainties. Future research including code amendments would be advisable for further understanding of the rich complex physics in this region.}

\subsection{Phase velocity of the instability and the Ion velocities}\label{App-PhaseVelocity}
Fig.~\ref{errorbars} shows that the ion velocities from the simulations (with the error bar from the above analysis) are very close to the phase velocities of the relativistic Buneman Instability (with error bars due to uncertainty in data extraction from simulation), implying a \textcolor{black}{possibility of} wave-particle Landau-resonance in ion acceleration.

\subsection{Stochasticity and ion energies}\label{Stoch_ions}
    \begin{figure}
    \centering
    \includegraphics[width=65mm]{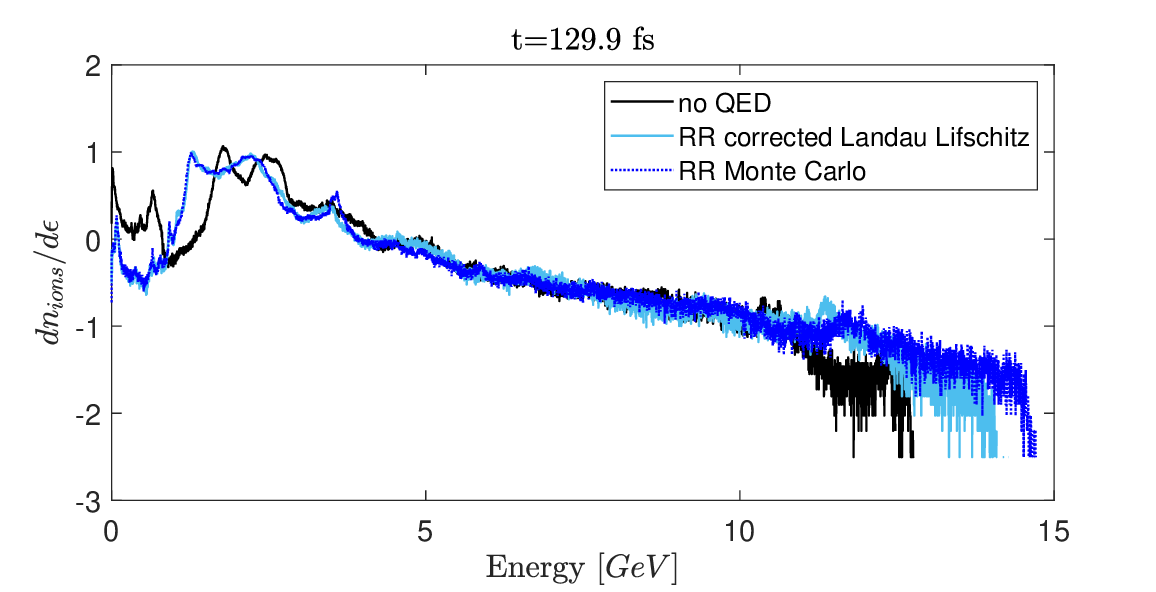}
    \caption{The ion energy spectra from SMILEI without RR (black), with RR using corrected Landau Lifschitz model that excludes stochasticity (sky blue), with more accurate Monte Carlo description (dark blue). }\label{spec}
    \end{figure}
\textcolor{black}{ Fig.~\ref{spec} shows ion energy spectra without RR (black), with RR using a corrected Landau-Lifschitz model (sky blue) and with a more accurate Monte Carlo description~\cite{Duclous_2011} (dark blue). The ion-energy enhancement is observed in both yet this is to different magnitudes where the Monte-Carlo model predicts a larger value of $E_{max}$.} \textcolor{black}{Energy-enhancement by stochastic effects captured by the Monte-Carlo method is also observed in Ref.~\cite{Wan.2019} where a circularly polarised laser is used to study RPA of ions.}  \textcolor{black}{From Fig.~\ref{pxpy} [panels (b) and (c)] we see that stochastic effects allow electrons to have larger longitudinal momentum (more than 800 $m_ec$ in panel (c)) even though the degree of collimation is not significantly different. This highligths the significance of high-energy driver-electrons in instability that facilitates a higher energy of ions with stochastic effects.}
  
%

\end{document}